\documentclass[twocolumn,showpacs,amsmath,amssymb]{revtex4}

\usepackage{graphicx,amssymb}
\usepackage{bm}

\newcommand \be{\begin{equation}}
\newcommand \en{\end{equation}}
\newcommand \bea{\begin{eqnarray}}
\newcommand \ena{\end{eqnarray}}

\begin{document}

\title{Asymptotically anomalous black hole configurations in gravitating nonlinear electrodynamics}

\author{J. Diaz-Alonso} \email{joaquin.diaz@obspm.fr}
\author{D. Rubiera-Garcia} \email{diego.rubiera-garcia@obspm.fr}
\affiliation{LUTH, Observatoire de Paris, CNRS, Universit\'e Paris
Diderot. 5 Place Jules Janssen, 92190 Meudon, France}
\affiliation{Departamento de F\'isica, Universidad de Oviedo. Avda.
Calvo Sotelo 18, E-33007 Oviedo, Asturias, Spain}

\date{\today}

\begin{abstract}

We analyze the class of non-linear electrodynamics minimally coupled to gravitation supporting asymptotically flat \textit{non Schwarzschild-like} elementary solutions. The Lagrangian densities governing the dynamics of these models in flat space are defined and fully characterized as a subclass of the set of functions of the
two standard field invariants, restricted by requirements of regularity, parity invariance and positivity of the energy, which are necessary conditions for the theories to be physically admissible. Such requirements allow for a complete characterization and classification of the geometrical structures of the elementary solutions for the corresponding gravity-coupled models. In particular, an immediate consequence of the requirement of positivity of the energy is the asymptotic flatness of gravitating elementary solutions for any admissible model. The present analysis, together with the (already published) one concerning the full class of admissible gravitating non-linear electrodynamics supporting asymptotically flat \textit{Schwarzschild-like} elementary solutions, completes and exhausts the study of the gravitating point-like charge problem for this kind of models.

\end{abstract}


\maketitle
\section{Introduction}

The study of gravitating non-linear electrodynamics (G-NED) generalizing the Reissner-Nordstr\"om (RN) solution of the Einstein-Maxwell field equations has received a considerable attention during the last decades, mainly due to the finding that (abelian and non-abelian) Born-Infeld (BI) models \cite{BI} arise, together with gravitation, in the low-energy regime of string and D-Brane physics \cite{string-BI}. In this context, asymptotically flat, static, spherically symmetric black hole solutions for the BI theory minimally coupled to gravity were obtained in a number of papers \cite{gravity-BI}. Moreover, other G-NED models supporting electrically charged black hole solutions have been obtained. As examples let us mention: the gravitating generalization of the Euler-Heisenberg effective Lagrangian of Quantum Electrodynamics \cite{gravity-EH}; the gravitating versions of a BI-like family of Lagrangians \cite{oliveira94} and of the logarithmic Lagrangian of Ref.\cite{gravity-log}; the black holes generated by Coulomb-like fields in (2+1) dimensions \cite{cataldo00}; those associated to models preserving the conformal invariance of Maxwell theory in any dimension \cite{hassaine07} as well as to models whose Lagrangian densities are defined as powers of the Maxwell Lagrangian \cite{hassaine08}.

Most of the examples considered in this context in the literature are particular cases of the full class of \textit{admissible} non-linear electrodynamics (NED) in flat space whose Lagrangians are defined as arbitrary functions of the two first-order field invariants. Admissibility amounts to the restriction of these functions by some conditions endorsing the physical consistency of the associated models (essentially, the positive definite character of the energy and the parity invariance, aside from some requirements for the proper definition and regularity of the Lagrangian functions and their elementary solutions). The subclass of these models supporting stable and finite-energy, electrostatic spherically symmetric (ESS) solutions (non-topological solitons) in flat space has been fully characterized and extensively analyzed in Refs.\cite{dr07-2} and \cite{dr09}. The ESS solutions of the models belonging to this subclass behave at the center ($r=0$) and asymptotically ($r \rightarrow \infty$) in such a way that the integral of energy is convergent. Next, in Ref.\cite{dr10}, we have extended the analysis to those admissible models supporting ESS solutions for which the integral of energy may diverge at the center. The main purpose in this reference was the study of the gravitating electrostatic spherically symmetric (G-ESS) solutions of the family of G-NEDs obtained from the minimal coupling of this class of admissible NEDs to gravity. The geometrical structure of the corresponding G-ESS solutions was exhaustively characterized in terms of the central and asymptotic behaviours of the ESS solutions (or, equivalently, in terms of the behaviours of the Lagrangian functions in vacuum and on the boundary of their domains of definition at vanishing magnetic field). Aside from naked singularities, extreme black holes or two-horizons black holes, as in the RN case, other gravitational configurations, such as extreme or non-extreme black points and single-horizon black holes, appeared for the models with ESS \textit{soliton} solutions in flat space. They are \emph{in all cases} asymptotically flat configurations with well defined Arnowitt-Deser-Misner (ADM) masses.

In order to exhaust the analysis of the geometrical structure of G-ESS solutions associated to the full class of admissible NEDs minimally coupled to gravity we must study (besides those considered in Ref.\cite{dr10}) the models supporting ESS solutions whose integral of energy diverges asymptotically. This is the main purpose of the present paper. As we shall see, these G-ESS solutions exhibit rich geometrical structures and, despite their unpleasant asymptotic behaviour in absence of gravity, they are always asymptotically flat, even though they approach flatness at large distances slower than the Schwarzschild field, and the ADM mass cannot be defined. We shall use the adjective ``anomalous" in calling this behaviour.

All the above gravitating solutions contain curvature singularities at the center, event though the metric may be finite there. This is a consequence of a non-existence theorem formulated for purely electrically charged solutions \cite{bronnikov-theorem}. Strictly speaking this theorem concerns the G-ESS solutions of NEDs with Maxwellian weak-field limit. Nevertheless, in the last section, we shall extend this theorem by proving the unavoidable presence of a curvature singularity at the center of the G-ESS solutions associated to any \textit{admissible} NED model minimally coupled to gravity. Such a theorem can be circumvented for other kinds of configurations, such as purely magnetically charged solutions \cite{magnetic-sol} or through a coupling between different structures \cite{regular}. Let us mention that some models containing singularity-free electrically charged black hole solutions, found in Ref.\cite{ab}, correspond to non-admissible Lagrangian densities which suffer ``branching" features as functions of their arguments, as explained in \cite{bronnikov00}.

In section II we summarize the main results on admissible NEDs in flat space, extending the classification of Refs.\cite{dr09} and \cite{dr10} in order to include the models supporting asymptotically energy-divergent ESS solutions. Section III is devoted to the outline of the G-NED problem in the ESS case and its resolution for the different types of models in terms of the boundary and vacuum behaviours of the NED Lagrangians in flat space, focusing on the detailed analysis of those leading to asymptotically anomalous G-ESS solutions. We conclude in section IV with a summary and some perspectives.

\section{Non-linear electrodynamics in flat space}

Let us recall the main results concerning the families of NEDs analyzed in references \cite{dr09} and \cite{dr10} and extend them to the full class of admissible NEDs. The generalized Lagrangian densities giving the dynamics of the fields are assumed to be general functions $\varphi(X,Y)$ of the field invariants

\bea
X &=& -\frac{1}{2} F_{\mu\nu}F^{\mu\nu} = \vec{E}^{2} - \vec{H}^{2}
\nonumber \\
Y &=& -\frac{1}{2} F_{\mu\nu}F^{*\mu\nu} = 2\vec{E}\cdot\vec{H},
\label{eq:(2-1)}
\ena
where the field strength tensor $F^{\mu\nu}$ and its dual $F^{*\mu\nu}$, as well as the electric $\vec{E}$ and magnetic $\vec{H}$ fields are defined in the usual way. These functions are restricted by some ``admissibility" conditions (such as regularity, positivity of the energy, parity invariance, etc.), which are necessary in order to define physically meaningful theories \cite{dr09}.

In terms of $\varphi(X,Y)$, the symmetric (gauge-invariant) energy-momentum tensor reads

\be
T_{\mu\nu} = 2F_{\mu\alpha}\left(\frac{\partial \varphi}{\partial X}F_{\nu}^{\alpha} + \frac{\partial \varphi}{\partial Y}F_{\nu}^{*\alpha}\right)- \varphi \eta_{\mu\nu},
\label{eq:(2-2)}
\en
and the energy density takes the form

\be
\rho = T_{00} = 2\frac{\partial \varphi}{\partial X}\vec{E}^{2} + 2\frac{\partial \varphi}{\partial Y} \vec{E}\cdot\vec{H} - \varphi(X,Y).
\label{eq:(2-3)}
\en
The necessary and sufficient condition for this energy functional to be positive definite and vanishing in vacuum, can be shown to be \cite{dr09}

\be
\rho \geq \left(\sqrt{X^{2}+Y^{2}} + X\right) \frac{\partial \varphi}{\partial X}+ Y\frac{\partial \varphi}{\partial Y} - \varphi(X,Y) \geq 0,
\label{eq:(2-4)}
\en
which requires $\varphi(0,0) = 0$, $\frac{\partial \varphi}{\partial X}\vert_{(X>0,Y=0)} > 0$ and $\varphi(X>0,0) > 0$. The Euler equations associated to a Lagrangian density $\varphi(X,Y)$ are

\be
\partial_{\mu} \left[\frac{\partial \varphi}{\partial X} F^{\mu\nu} + \frac{\partial \varphi}{\partial Y} F^{*\mu\nu}\right] = 0,
\label{eq:(2-5)}
\en
and the search for ESS solutions of the form $\vec{E}(\vec{r}) = E(r) \frac{\vec{r}}{r}, \vec{H} = 0$ lead to the first-integral

\be
r^{2} E(r) \frac{\partial \varphi}{\partial X} \Big \vert_{Y=0} = Q,
\label{eq:(2-6)}
\en
where $Q$ is an integration constant, identified as the source electric charge. Owing to the admissibility condition $\varphi_X \vert_{Y=0} >0$ the signs of $E(r)$ and $Q$ are the same and, consequently, we can consider only the case $Q>0$ without loss of generality \cite{dr09}. The solutions with $Q<0$ are straightforwardly obtained through the replacement $E(r,Q,\mu) \rightarrow sign(Q)E(r,\vert Q \vert, \mu)$. The form of the ESS solution of charge $Q$ obtained from Eq.(\ref{eq:(2-6)}) (with $E(r,Q) = \sqrt{X}$; see Eqs.(\ref{eq:(2-1)})) is completely specified once the explicit expression of $\varphi(X,0)$ is given. Owing to the admissibility conditions, $E(r,Q)$ is a monotonic function of $r$ \cite{dr09}. From Eq.(\ref{eq:(2-3)}) the energy density associated to these solutions reads

\be
\rho = 2\frac{\partial \varphi}{\partial X} \Big \vert_{Y=0} E^{2}(r,Q) - \varphi(E^{2}(r,Q),Y=0),
\label{eq:(2-7)}
\en
whose integral in space, if finite, gives the total energy as a function of the charge. Obviously, the finite or divergent-energy character of the ESS solutions for a given model (aside from its linear stability) is related to the form of the Lagrangian density $\varphi(X,Y)$ governing the dynamics. The complete characterization of the families of admissible NEDs supporting ESS non-topological soliton solutions in flat space has been performed in Refs. \cite{dr07-2,dr09}. Besides these models, we are interested here in the whole family of admissible NEDs. As we shall see, when minimally coupled to gravity, they lead to G-ESS solutions which are asymptotically flat space-times. Let us consider the behaviour of the ESS field around the center and as $r \rightarrow \infty$, assumed to be of the form \footnote{Although this power-law expression form excludes more complex transcendent behaviours at the center and asymptotically, the validity of our conclusions can be easily shown for these cases.}

\be
 E(r \rightarrow 0) \sim r^{p},
\label{eq:(2-8)}
\en
and

\be
 E(r \rightarrow \infty) \sim r^{q}.
\label{eq:(2-9)}
\en
Using Eqs.(\ref{eq:(2-3)}) and (\ref{eq:(2-4)}) we see that the \textit{positivity of the energy requires} $p \leq 0$ and $q < 0$. When $p < 0$ in (\ref{eq:(2-8)}) the ESS fields diverge at $r=0$, whereas for $p = 0$ the field must behave around the center as

\be
E(r \rightarrow 0) \sim a - b r^{\sigma},
\label{eq:(2-10)}
\en
where the maximum field strength ($a$) and the exponent ($\sigma > 0$) are characteristic parameters of the model, while the coefficient $b$ is related to the charge of the particular ESS solution (the quantity $\alpha = Q b^{2/\sigma}$ turns out to be a universal constant for a given model \cite{dr10}). Consequently, the solutions $\vert E(r,Q) \vert$ of the admissible models decrease monotonically from $+ \infty$ (if $p < 0$), or from a finite value $\vert a \vert$ (if $p = 0$), at $r = 0$, and vanish asymptotically as $r \rightarrow \infty$. The behaviour of the energy density around the center is given by

\be
\rho(r \rightarrow 0) \sim r^{p-2},
\label{eq:(2-11)}
\en
(see Eqs.(\ref{eq:(2-6)}) and (\ref{eq:(2-7)})) so that the integral of energy converges there if $-1 < p \leq 0$ and diverges if $p \leq -1$. Following our previous conventions \cite{dr09} we denote class-\textbf{A1} models those for which the ESS solutions behave as in Eq.(\ref{eq:(2-8)}) with the exponent $-1 < p < 0$ and class-\textbf{A2} models those supporting ESS solutions which behave as in Eq.(\ref{eq:(2-10)}) (corresponding to $p = 0$ in Eq.(\ref{eq:(2-8)})). In both cases the integral of energy converges at the center. On the other hand we shall call ``ultraviolet divergent" (\textbf{UVD}) those models with $p \leq -1$, for which the integral of energy of the ESS solutions diverges at the center.

Asymptotically, the energy density behaves as

\be
\rho(r \rightarrow \infty) \sim r^{q-2},
\label{eq:(2-12)}
\en
and the integral of energy converges there if $q < -1$, but diverges for $-1 \leq q < 0$. In our previous work the asymptotically energy-convergent cases were split into three subclasses: \textbf{B1} models with $-2 < q < -1$ (ESS solutions damped slower than the coulombian field); \textbf{B2} models with $q = -2$ (coulombian damping); \textbf{B3} models with $q < -2$ (faster than coulombian damping). For models whose ESS solutions are damped with an exponent $-1 \leq q < 0$ in Eq.(\ref{eq:(2-9)}), the integral of energy diverges asymptotically. We shall call them \textbf{IRD} models (acronym for ``infrared divergent").

Let us analyze now the behaviour of the Lagrangian densities around the values of their arguments corresponding to the central and asymptotic regions of the ESS solutions (see Fig.1). From the first-integral (\ref{eq:(2-6)}) we see that the Lagrangian density behaves, in the A1 and UVD cases, as

\be
\varphi(X,Y=0) \sim X^{\gamma},
\label{eq:(2-12)bis}
\en
around $X = E^{2}(r \rightarrow 0) \rightarrow \infty$, where

\be
\gamma = \frac{p-2}{2p}.
\label{eq:(2-12)ter}
\en
The condition (\ref{eq:(2-4)}) for the positivity of the energy reads in these cases

\be
(2\gamma - 1) X^{\gamma} \geq 0,
\label{eq:(2-12)quart}
\en
and this equation confirms that if $p > 0$ (so that the exponent $\gamma < 1/2$) the energy density becomes negative around the vacuum. The A1 cases correspond to $\gamma > 3/2$ whereas the UVD cases correspond to $1/2 < \gamma \leq 3/2$. For $p = 0$ (case A2) the field at the center behaves as in Eq.(\ref{eq:(2-10)}) and the Lagrangian density behaves around $(X = E^{2}(r=0) = a^{2}, Y=0)$ as

\be
\varphi(X,Y=0) \sim \frac{2\alpha \sigma}{2-\sigma}(a - \sqrt{X})^{\frac{\sigma-2}{\sigma}} + \Delta,
\label{eq:(2-12)q}
\en
if $\sigma \neq 2$. Here $\alpha$ and $\Delta$ are universal constants of the model \cite{dr10}. If $\sigma = 2$ we have

\be
\varphi(X,Y=0) \sim -\alpha \ln(a - \sqrt{X}).
\label{eq:(2-12)six}
\en
Consequently, for $\sigma \leq 2$ the Lagrangians exhibit a vertical asymptote at this point. On the other hand they take a finite value there with infinite slope for $\sigma > 2$ (the BI model belongs to this case with $\sigma = 4$).

The same equation (\ref{eq:(2-12)bis}), now with

\be
\gamma = \frac{q-2}{2q},
\label{eq:(2-12)s}
\en
describes the behaviour of the Lagrangian density around the vacuum $(X = E^{2}(r \rightarrow \infty) = 0, Y=0)$ and the ranges of values of the exponent are now $1 < \gamma < 3/2$ in case B1, $\gamma = 1$ in case B2, $1/2 < \gamma < 1$ in case B3 and $\gamma \geq 3/2$ in case IRD. All these behaviours of the admissible NEDs are plotted in figure 1.

\begin{figure}
\includegraphics[width=8.6cm,height=6.5cm]{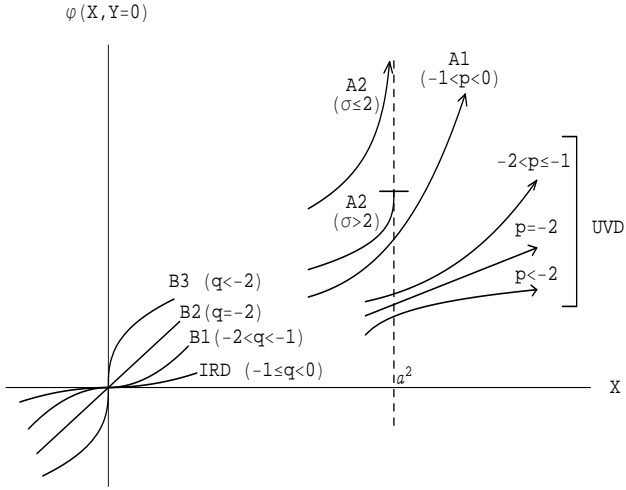}
\caption{\label{fig:1} Behaviours of the Lagrangian densities of admissible NEDs around the vacuum (cases B1, B2, B3 and IRD) and on the boundary of their domain of definition at $Y = 0$ (cases A1, A2 and UVD). In between both domains, these functions must be strictly monotonic increasing in order for the energy to be positive definite. The parameters $p$, $\sigma$ and $q$ correspond to the exponents of $r$ in Eqs.(\ref{eq:(2-8)}), (\ref{eq:(2-10)}) and (\ref{eq:(2-9)}), respectively, determining the central and asymptotic field behaviours of the ESS solutions.}
\end{figure}

We see from this discussion and the regularity of $E(r>0,Q)$ that we can define, in cases A1 and A2, the \textbf{internal-energy function}

\be
\varepsilon_{in}(r,Q) = 4\pi \int_{0}^{r} R^{2} \rho(R,Q) dR,
\label{eq:(2-13)}
\en
giving the energy of an ESS solution of charge $Q$ inside a sphere of radius $r$. Owing to the admissibility conditions, this is a \emph{monotonically increasing and convex} function of $r$. This function cannot be defined in the UVD cases. In the asymptotic cases B1, B2 and B3 we can define the \textbf{external-energy function}

\be
\varepsilon_{ex}(r,Q) = 4\pi \int_{r}^{\infty} R^{2} \rho(R,Q) dR,
\label{eq:(2-14)}
\en
which gives the energy of the field outside the sphere of radius $r$. This is a \emph{monotonically decreasing and concave} function \cite{dr10}, which cannot be defined in the IRD cases.

The six combinations of cases A1 and A2 for the central-field ESS behaviour with cases B1, B2 and B3 for the asymptotic behaviour, lead to six families of models which exhaust the class of admissible NEDs supporting finite-energy ESS solutions. In such theories $\varepsilon_{in}(r,Q)$ and $\varepsilon_{ex}(r,Q)$ are well defined and satisfy

\be
\varepsilon(Q) = \varepsilon_{in}(\infty,Q) = \varepsilon_{ex}(0,Q) = \varepsilon_{in}(r,Q) + \varepsilon_{ex}(r,Q),
\label{eq:(2-15)}
\en
where $\varepsilon(Q)$ is the total energy of the ESS solution of charge $Q$, scaling as $\varepsilon(Q) = Q^{3/2}\varepsilon(Q=1)$.

The \textit{full set} of admissible non-linear electromagnetic field theories is now obtained by including, besides the soliton-supporting ones, the families of admissible models whose associated ESS solutions are energy-divergent, exhibiting UVD, IRD or both behaviors. This extended set is now naturally classified into twelve families, by combining the three central-field behaviors (A1, A2 and UVD) with the four asymptotic behaviors (B1, B2, B3 and IRD).

\section{Gravitating non-linear electrodynamics}

Let us consider now the minimal coupling of admissible NEDs to the gravitational field, defined through the action

\be
S=S_{G} + S_{NED} = \int d^4x \sqrt{-g}\left[\frac{R}{16\pi G} - \varphi(X,Y)\right],
\label{eq:(3-1)}
\en
where, as usually, $g$ and $R$ are the metric determinant and the curvature scalar, respectively. The variation of this action with respect to the electromagnetic fields lead to field equations which are obtained from Eqs.(\ref{eq:(2-5)}) by the replacement of partial derivatives by covariant derivatives. The expression of the mixed components of the energy-momentum tensor of the electromagnetic fields, from which the Einstein equations can be immediately written, remain the same as in flat space in the static spherically symmetric case, where these equations, together with the relations between the components of the energy-momentum tensor

\be
T_{0}^{0} = T_{1}^{1} = 2E^{2} \frac{\partial \varphi}{\partial X} \Big \vert_{Y=0} - \varphi \hspace{.25cm};\hspace{.25cm} T_{2}^{2} = T_{3}^{3} = -\varphi,
\label{eq:(3-1)bis}
\en
allow for the introduction of an adapted coordinate system where the interval takes the Schwarzschild-like form \cite{dr10}

\be
ds^{2} = \lambda(r) dt^{2} - \frac{dr^{2}}{\lambda(r)} - r^{2} d\Omega^{2},
\label{eq:(3-2)}
\en
with $d\Omega^{2} = d\theta^{2} + \sin^{2} \theta d\phi^{2}$. The field equations for the G-ESS solutions in this coordinate system take the same form as those in spherical coordinates in the flat-space problem for the ESS solutions. Consequently, the first-integral (\ref{eq:(2-6)}) and the form of the ESS field solutions, as functions of the radial coordinates $r$, are the same in both cases. Thus, the analysis and classification of the ESS solutions of the admissible models in flat space in terms of their central and asymptotic behaviours, can be immediately translated to the gravitating problem.

On the other hand, the Einstein equations in this G-ESS case take the form

\bea
\frac{d}{dr}\left(r\lambda(r) - r\right) &=& -8\pi r^{2} T^{0}_{0} = -8\pi r^{2}\left(2\frac{\partial \varphi}{\partial X} E^{2} - \varphi\right)\nonumber \\
\frac{d^{2}}{dr^{2}}\left(r\lambda(r)\right) &=& -16\pi r T^{2}_{2} = 16\pi r \varphi,
\label{eq:(3-3)}
\ena
whose compatibility can be easily established. The general solution of these equations reads

\be
\lambda(r,Q,C) = 1 + \frac{C}{r} - \frac{8\pi}{r}\int r^{2} T^{0}_{0}(r,Q) dr,
\label{eq:(3-4)}
\en
where $C$ is an arbitrary constant, absorbing the constant of the indefinite integral of the energy density term in this formula. When the asymptotic behaviour of the ESS solutions belongs to cases B1, B2 or B3, Eq.(\ref{eq:(3-4)}) can be written as

\be
\lambda(r,Q,M) = 1 - \frac{2M}{r} + \frac{2\varepsilon_{ex}(r,Q)}{r},
\label{eq:(3-5)}
\en
where $\varepsilon_{ex}(r,Q)$ is the external-energy function in flat space, defined in Eq.(\ref{eq:(2-14)}), and the constant $M$ is identified as the ADM mass. The structure of the G-ESS solutions of these admissible NEDs minimally coupled to gravity has been classified and extensively analyzed in Ref.\cite{dr10}. This analysis concerned the models obtained from the UVD, A1 and A2 central field behaviours combined with the B1, B2 and B3 asymptotic behaviours. The corresponding metrics are asymptotically flat, behaving for large $r$ as the Schwarzschild one, no matter the central-field behaviour. If this central behaviour is UVD, the structure of the metrics (\ref{eq:(3-5)}) for different values of $M$ and $Q$ is qualitatively the same as that of the Reissner-Nordstr\"om solution of the Einstein-Maxwell field equations (naked singularities, extreme black holes or two-horizons black holes). If the central behaviour belongs to cases A1 or A2 we have, in addition, new kinds of solutions (non-extreme single-horizon black holes and extreme and non-extreme black points). However, as already mentioned in the introduction, the analysis of admissible models supporting IRD solutions in flat space, which lead also to asymptotically flat G-ESS configurations, was not considered in that reference. We shall now tackle this issue by exploring the properties of such solutions. These admissible models support ESS solutions in flat space behaving asymptotically as in Eq.(\ref{eq:(2-9)}) with $0 > q \geq -1$ and, as a consequence of the asymptotic behaviour of the energy density (\ref{eq:(2-12)}), the integral of energy diverges at large $r$. Consequently, $\varepsilon_{ex}(r,Q)$ in (\ref{eq:(2-14)}) cannot be defined. We are thus lead to three new cases by combining this asymptotic behaviour with the A1, A2 or UVD central-field behaviours, for which the methods of Ref.\cite{dr10} cannot be immediately applied. Let us analyze these cases separately.

\begin{figure}
\begin{center}
\includegraphics[width=8.6cm,height=6cm]{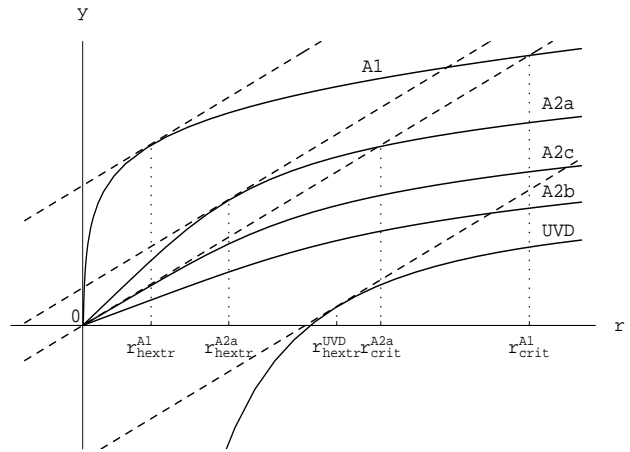}
\caption{\label{fig:2} Qualitative behaviour of the energy curves $y=\varepsilon_{in}(r,Q)$ (for fixed $Q$)
for the different admissible IRD models (the curve UVD corresponds to the primitive of $r^{2}T_{0}^{0}(r,Q)$,
defined up to an arbitrary constant). The slopes of these curves at $r=0$ diverge in the A1 and UVD cases and
take finite values ($= 8\pi Q a$) in the A2 cases. In all cases these curves are monotonic, convex and diverge
asymptotically with vanishing slope. The cutting points between these curves and the beam of dashed lines
$y = \frac{r+C}{2}$ define the horizons of the configurations. The tangency points correspond to extreme black hole
and extreme black point configurations. Otherwise the configurations are naked singularities and one or two-horizons
black holes. The cutting points between the energy curves and the straight line $C=0$ provide the radius of the critical
configurations (which are meaningless for the UVD case).}
\end{center}
\end{figure}

\subsection{IRD-A1 case}

The exponents in Eqs.(\ref{eq:(2-8)}) and (\ref{eq:(2-9)}) range now in the intervals $-1 < p < 0$ and $-1\leq q <0$, respectively. The integration of the Einstein equations (\ref{eq:(3-3)}) leads to

\be
\lambda(r,Q,C) = 1 + \frac{C}{r} - \frac{2\varepsilon_{in}(r,Q)}{r},
\label{eq:(3-6)}
\en
where $\varepsilon_{in}(r,Q)$ is the interior integral of energy in flat space, defined by Eq.(\ref{eq:(2-13)}) and $C$ is an arbitrary integration constant. Owing to equation (\ref{eq:(2-12)}), $\varepsilon_{in}(r,Q)$ diverges asymptotically as

\be
\varepsilon_{in}(r \rightarrow \infty,Q) \sim r^{q+1} \rightarrow \infty,
\label{eq:(3-7)}
\en
exhibiting a horizontal parabolic branch. Thus the last term in the r.h.s. of (\ref{eq:(3-6)}) dominates over the second one for large $r$ and the metric approaches asymptotic flatness as $1 - \lambda(r \rightarrow \infty) \sim r^{q}$, slower than the Schwarzschild field. Consequently, there is not a well defined mass for these configurations (in fact, the expression for the ADM mass \cite{ortin} diverges for all the IRD cases). At $r=0$ equations (\ref{eq:(2-11)}) and (\ref{eq:(2-13)}) show that $\varepsilon_{in}(r=0,Q) = 0$ and this function exhibits a vertical slope there (see figure 2). The term $C/r$ in (\ref{eq:(3-6)}) dominates at small $r$ and the metric function diverges at the center with the sign of $C$ (see figure 3). The condition $\lambda(r,Q,C) = 0$, determining the radii of the horizons, leads to

\be
\varepsilon_{in}(r,Q) = \frac{r + C}{2},
\label{eq:(3-8)}
\en
which relates these radii and the values of $C$ for a given charge. Consequently, the horizon locations are given by the intersection points between the beam of straight lines $y = \frac{r + C}{2}$ and the monotonically increasing function $y = \varepsilon_{in}(r,Q)$. The straight line of the beam which is tangent to the internal-energy function defines an extreme black hole configuration, which is determined from Eq.(\ref{eq:(3-8)}), aside from the condition

\be
\frac{\partial \varepsilon_{in}}{\partial r} = 4\pi r^{2} T^{0}_{0}(r,Q) = \frac{1}{2}.
\label{eq:(3-9)}
\en
This condition gives the tangency point location $r_{hextr}(Q)$, which is the horizon radius of the extreme black hole configuration and, in addition, Eq.(\ref{eq:(3-8)}) gives the value of the constant $C_{extr}(Q)$ of this tangent line, both parameters being functions of the charge. There are now several possibilities, depending on the values of $C$ as compared to the one of $C_{extr}(Q)$ (see figures 2 and 3):

1) \textbf{$C > C_{extr}(Q)$}: There are no horizons and the solutions exhibit \textbf{naked singularities} at the center, where the metric function $\lambda(r,Q)$ diverges to $+\infty$ (line I on Fig.3).

2) \textbf{$C = C_{extr}(Q)$}: The solutions are \textbf{extreme black holes} exhibiting a degenerate horizon at $r_{hextr}(Q)$ (line II on Fig.3).

3) \textbf{$0 < C < C_{extr}(Q)$}: The solutions are \textbf{two-horizons black holes}. As $C$ decreases in this range, the inner (Cauchy) horizon radius decreases from $r_{hextr}(Q)$ to zero, whereas the outer (event) horizon radius increases from $r_{hextr}(Q)$ to a value $r_{hcrit}(Q)$ (lines III and IV on Fig.3). For the limit of these solutions obtained as $C \rightarrow 0^{+}$ the metric function $\lambda(r,Q, C \rightarrow 0^{+})$ diverges at $r = 0$, stepping from $+\infty$ to $-\infty$ in crossing a vanishing-radius inner horizon.

4) \textbf{$C = 0$}: Consistently with the conventions of Ref.\cite{dr10} we shall call this one the \emph{critical configuration}. The metric function coincides with that of the limit of the solutions of the precedent case 3) for $r > 0$, but diverges to $-\infty$ at $r = 0$ (no step at the center). There is a unique horizon at $r = r_{hcrit}(Q)$ and the configuration is a \textbf{single-horizon black hole} (dashed line V in Fig.3).

5) \textbf{$C < 0$}: The configurations are \textbf{single-horizon black holes} and $\lambda(r,Q,C)$ increases monotonically from $-\infty$ at $r = 0$ to $\lambda \rightarrow 1$ as $r \rightarrow \infty$ (line VI on Fig.3). The limit of these solutions obtained as $C \rightarrow 0^{-}$ coincides with the critical solution of the precedent case for any $r$ and, for $r > 0$, with the limit solution of case 3) above.

The causal nature of the central curvature singularities of the 1) to 3) configurations ($C > 0$) is timelike whereas for both 4) and 5) configurations ($C \leq 0$) these singularities are spacelike. As mentioned above, the metric function of the limit configuration obtained as $C \rightarrow 0^{+}$, coincides with the one of the critical configuration for $r > 0$, but differs by an infinite step at the center. This step introduces supplementary $\delta$-like distribution terms in the curvature tensor which modify the causal structure of the singularity \cite{ortin}.

\begin{figure}
\begin{center}
\includegraphics[width=8.6cm,height=6cm]{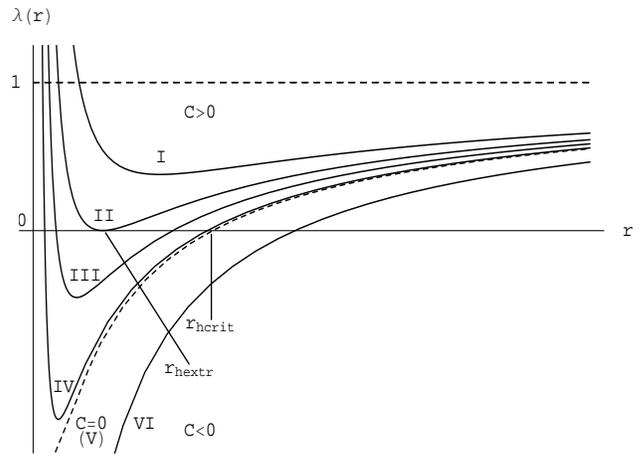}
\caption{\label{fig:3} Behaviour of $\lambda(r)$ for the IRD-A1 configurations. When $C>0$ there can be
naked singularities (curve I, $C > C_{extr}(Q)$), extreme black holes (curve II, $C = C_{extr}(Q))$ or two-horizons
black holes (curves III and IV, $0 < C < C_{extr}(Q)$). Solutions with $C \leq 0$ lead to single-horizon black holes
(curve V, $C = 0$, corresponding to the critical configuration and curve VI for $C < 0$). The structure of the limit
configurations obtained as $C \rightarrow 0^{\pm}$ can be easily visualized in this figure. All curves approach
asymptotic flatness slower than the Schwarzschild solution. Curves I to IV describe also the typical qualitative
behaviour of the metric function in the IRD-UVD cases.}
\end{center}
\end{figure}

As a simple example illustrating these IRD-A1 models let us look for a family of NEDs supporting ESS solutions of the form

\be
E(r,Q,s) = \frac{Q^{s/2}}{r^{s}},
\label{eq:(3-10)}
\en
where $0 < s < 1$ parameterizes the family and $Q$ is the integration constant characterizing the solution, which will be identified as the electric charge. It is introduced under this form because the ESS solutions must depend on the radius through the ratio $r/\sqrt{Q}$, as results from the first-integral (\ref{eq:(2-6)}) (see Ref.\cite{dr09}). By using this first-integral we are easily lead to a family of NED Lagrangians $\varphi(X,Y,s)$ behaving on the $Y = 0$ axis as \footnote{Aside from this condition the Lagrangian may be continued to other regions of the $X-Y$ plane in such a way that the admissibility conditions be satisfied (see Ref.\cite{dr09}, section 3). Obviously, such an extension is not unique. The G-ESS solutions of a subclass of this family, defined by Lagrangian densities built as integer powers of the Maxwell Lagrangian in $d$-dimensions, have been considered in reference \cite{hassaine07}.}

\be
\varphi(X,Y=0,s) = X^{\gamma},
\label{eq:(3-11)}
\en
for $X>0$, and where the parameter $\gamma = \frac{s+2}{2s}$, characterizing the different models in the family, is restricted to the range $3/2 < \gamma < \infty$. By integrating (\ref{eq:(2-7)}) in this case, the internal-energy function takes the form

\be
\varepsilon_{in}(r,Q,s) = 8\pi \Theta(Q,s) r^{1-s},
\label{eq:(3-12)}
\en
where

\be
\Theta(Q,s) = \frac{1}{s(1-s)} \left(\frac{2sQ}{s+2}\right)^{(1+s/2)}.
\label{eq:(3-13)}
\en
As expected this is a monotonically increasing and convex function of $r$, with a divergent slope at the origin and diverging with vanishing slope as $r \rightarrow \infty$. The metric function reads

\be
\lambda(r,C,Q,s) = 1 + \frac{C}{r} - 16\pi \frac{\Theta(Q,s)}{r^{s}},
\label{eq:(3-14)}
\en
and defines a beam of curves having the form shown in Fig.3. In all cases these metrics approach asymptotic flatness slower than Schwarzschild or RN configurations. The radii of the horizons and the values of the constants $C(Q)$ for extreme black hole solutions are obtained from Eqs.(\ref{eq:(3-8)}) and (\ref{eq:(3-9)}), respectively. The horizon radii of the critical black hole configurations result from the condition $\lambda(r,C=0,Q,s) = 0$. These values are

\bea
r_{hextr}(Q,s) &=& (32 \pi (1-s) \Theta(Q,s))^{1/s} \nonumber \\
C_{extr}(Q,s) &=& \frac{s}{1-s} r_{hextr}(Q,s)\\
r_{hcrit}(Q,s) &=& \frac{r_{hextr}(Q,s)}{(1-s)^{1/s}} > r_{hextr}(Q,s). \nonumber
\label{eq:(3-15)}
\ena
The verification for this family of the remaining results 1)-5) established for the general IRD-A1 models is now straightforward.

\subsection{IRD-A2 case}

In this case the expression of the metric function is still given by Eq.(\ref{eq:(3-6)}), but $\varepsilon_{in}(r,Q)$ has now a finite slope at the center given by

\be
\frac{\partial \varepsilon_{in}}{\partial r} \Big \vert_{r=0} = 4 \pi r^{2} T_{0}^{0} \Big \vert_{r=0} = 8\pi Q a,
\label{eq:(3-16)}
\en
as results from Eqs.(\ref{eq:(2-7)}) and (\ref{eq:(2-11)}) (see Fig.2). Consequently, the internal-energy function $\varepsilon_{in}(r,Q)$ takes a form which is similar to that of the precedent case IRD-A1, excepting that the slope at the center is now finite. In this case the metric function and its derivative behave near the center as

\bea \label{eq:(3-17)}
\lambda(r \rightarrow 0,C,Q,a,b,\sigma) &\sim& 1 - 16\pi Qa + \frac{C}{r} + \nonumber \\ &+& \frac{32\pi bQ}{(\sigma+1)(2-\sigma)}r^{\sigma}+ \Delta r^{2}\\
\frac{\partial \lambda}{\partial r} \Big \vert_{r \rightarrow 0} \sim -\frac{C}{r^{2}} &+& \frac{32\pi b Q \sigma}{(\sigma+1)(2-\sigma)} r^{\sigma-1}+ 2\Delta r, \nonumber
\ena
for $\sigma \neq 2$ and

\bea \label{eq:(3-18)}
\lambda(r \rightarrow 0,C,Q,a,b) &\sim& 1 - 16\pi Qa + \frac{C}{r} + \nonumber \\ &+& \frac{8\pi b Q}{3} r^{2}\left(1 - 2 \ln(r)\right)+ \Delta r^{2}  \\
\frac{\partial \lambda}{\partial r} \Big \vert_{r \rightarrow 0} &\sim& -\frac{C}{r^{2}} - \frac{32\pi b Q}{3} r \ln(r)+ 2\Delta r, \nonumber
\ena
for $\sigma = 2$, $\Delta$ being an integration constant. When $\sigma > 2$ this constant is given by $\Delta = \frac{8\pi Q}{3} \varphi(a^{2},0)$. Otherwise the value of $\Delta$ is not relevant because the associated term is not dominant in these equations. We must now distinguish three subcases, according to $16 \pi Q a \lesseqqgtr 1$ (see figure 2).

a) For the first subcase ($16 \pi Q a < 1$; curve A2b in Fig.2) the beam of straight lines with $C < 0$ cut the curve $\varepsilon_{in}(r,Q)$ once, leading to \textbf{single-horizon black hole} configurations. When $C \geq 0$ the configurations are timelike \textbf{naked singularities} for which the metric function diverges at the center if $C > 0$, and attains a finite positive value there ($\lambda(0) = 1 - 16\pi Qa > 0$) for the critical configuration ($C = 0$). For $C < 0$ the configurations have a spacelike curvature singularity at the center, where the metric function is negatively divergent. The limit configuration ($C \rightarrow 0^{-}$) is a \textbf{black point}, for which the metric function steps from $-\infty$ at the center to $\lambda(r  \rightarrow 0^{+})  \rightarrow 1 - 16\pi Qa > 0$ on the outer side of the vanishing-radius horizon, coinciding with the critical configuration for $r > 0$. The limit configuration ($C \rightarrow 0^{+}$) is a \textbf{naked singularity}, exhibiting also a singularity of the metric function $\lambda(r)$ at the center, which steps there from $+\infty$ to the finite value $1 - 16 \pi Q a $. For $r > 0$ this limit configuration coincides with the critical one and with the limit black point one. As easily seen from Eqs.(\ref{eq:(3-17)}) and (\ref{eq:(3-18)}) the slope of the critical metric function at the center is related to the value of the $\sigma$ parameter of the model, diverging for $\sigma < 1$ (see Fig.4), vanishing for $\sigma > 1$ (see Fig.5) and taking a finite value for $\sigma = 1$ (see Fig.6).

b) For the second subcase ($16 \pi Q a > 1$; curve A2a in Fig.2) there is a tangency point of one of the straight lines of the beam $\frac{r + C}{2}$, defining \textbf{extreme black hole} configurations, where $C = C_{extr}(Q)$ and the extreme horizon radius $r_{hextr}(Q)$ are given by equations (\ref{eq:(3-8)}) and (\ref{eq:(3-9)}). For different values of the constant $C$, corresponding to the different straight lines of the above beam, the configurations are similar to those of the IRD-A1 case ($C > C_{extr}(Q)$: \textbf{naked singularities}; $0 < C < C_{extr}(Q)$: \textbf{two-horizons black holes}; $C \leq 0$: \textbf{single-horizon black holes}). The causal structure of the central singularity is timelike in the cases $C > 0 (\lambda(r \rightarrow 0) \rightarrow +\infty)$ and spacelike in the cases $C < 0 (\lambda(r \rightarrow 0) \rightarrow -\infty)$. For the critical configuration ($C = 0$) the metric function in Eqs.(\ref{eq:(3-17)}) and (\ref{eq:(3-18)}) attains a negative finite value at the center, given by $\lambda(r=0,Q) = 1 - 16\pi Q a < 0$ and, consequently, the causal structure of the central singularity is spacelike. For the limit configurations obtained as $C \rightarrow 0^{\pm}$ the metric functions step at the center from $\pm \infty$, respectively, to this value, coinciding with the critical configuration for $r > 0$. Thus, the causal structure of the central singularities of these limit configurations is timelike and spacelike, respectively. As in the precedent subcase, the slope of $\lambda(r)$ at the center for the critical configuration depends on the value of the model parameter $\sigma$ in the same way (see Figs.4, 5 and 6).

c) For the third subcase ($16 \pi Q a = 1$; curve A2c in Fig.2) the straight line of the beam corresponding to $C = 0$ is tangent to the internal-energy curve at $r = 0$. The corresponding critical configuration is an \textbf{extreme black point}. The metric function for this configuration vanishes at the center and its slope there depends on the value of the parameter $\sigma$, as in the precedent subcases. The configurations with $C > 0$ are \textbf{naked singularities}, whereas those with $C < 0$ are \textbf{single-horizon black holes}. The limit configuration obtained as $C \rightarrow 0^{+}$ is a \textbf{black point} with a discontinuity of the metric function, which steps from $+\infty$ at the center to zero on a vanishing-radius emerging horizon and coincides with the critical one for $r > 0$. The limit configuration obtained as $C \rightarrow 0^{-}$ is a \textbf{black point} for which the metric function steps from $-\infty$ to zero in crossing the vanishing radius horizon, coinciding also for $r > 0$ with the critical one. There is a null curvature singularity at the center of the extreme black point critical configuration, but the singularities of the black point limit configurations obtained as $C \rightarrow 0^{\pm}$ are timelike and spacelike, respectively.

\begin{figure}
\begin{center}
\includegraphics[width=8.6cm,height=6cm]{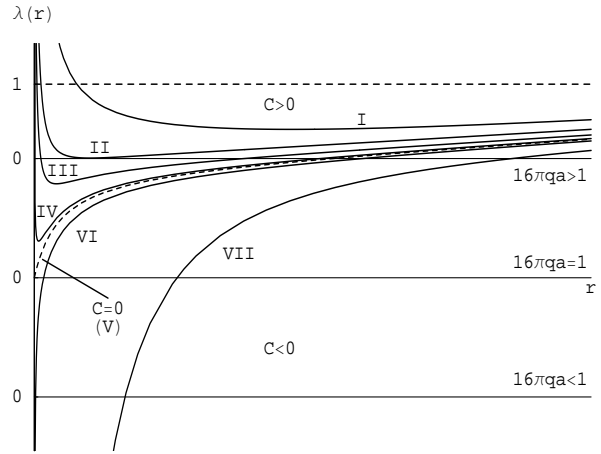}
\caption{\label{fig:4} Qualitative behaviour
of $\lambda(r)$ for the IRD-A2 models when $\underline{\sigma >1}$. This behaviour is similar in the three cases
\emph{b), c), a)}, ($16\pi qa \gtreqqless 1$, respectively), excepting by the sign of the metric function of the critical
configurations at the center, which is $\lambda_{crit}(r = 0) \lesseqqgtr 0$, respectively. Then, by simplicity,
we have plotted a unique set of curves for the different configurations and modified the relative position of the
$\lambda = 0$ axis for each case (continuous horizontal lines). The metric functions diverge at the center to
$+\infty$ when $C>0$ and to $-\infty$ for $C<0$, whereas for the critical configuration ($C = 0$), they attain a
finite value, which is positive, null or negative, depending on the case. The slope of the critical configuration
at the center diverges for this value of $\sigma$. Curve I corresponds to naked singularities in all cases. Curve
II corresponds to extreme black hole configurations in the $16\pi qa > 1$ case and to naked singularities otherwise. Curves III and IV correspond to two-horizons black holes in the $16\pi qa > 1$ case and to naked singularities
otherwise. The dashed curve V is the critical configuration and corresponds to single-horizon black holes in
the $16\pi qa > 1$ case, to extreme black points when $16\pi qa = 1$ and to naked singularities in the
$16\pi qa < 1$ case. Curves VI and VII correspond to single-horizon black holes in all cases. Curves IV and VI
show how the metric approaches the behaviour of the critical case as $C \rightarrow 0$ from above and below, respectively.}
\end{center}
\end{figure}

\begin{figure}
\begin{center}
\includegraphics[width=8.6cm,height=6cm]{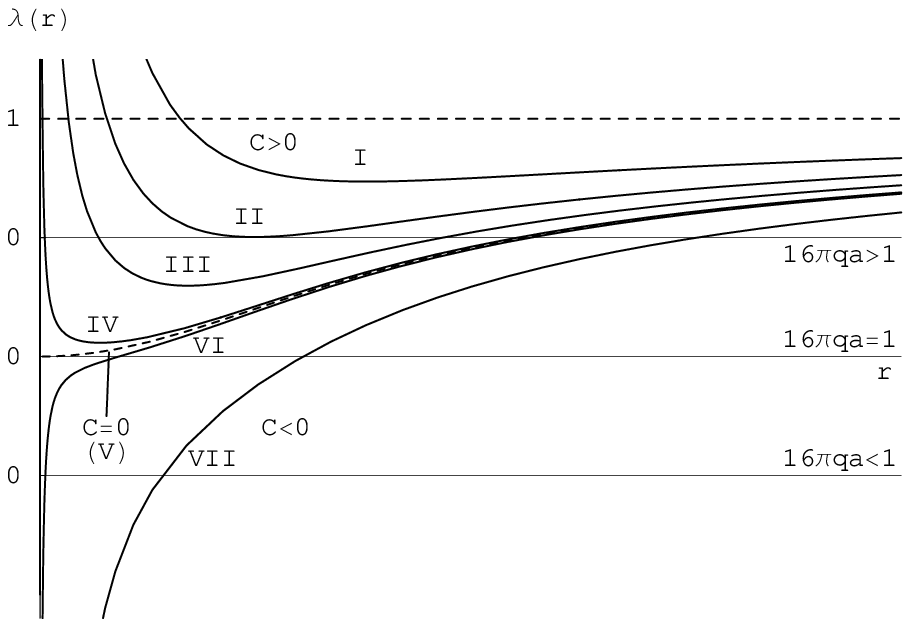}
\caption{\label{fig:5} Same comments as in Fig.4, but now for \underline{$\sigma < 1$}. The slope at the center
of the metric function for the critical configuration ($C=0$) vanishes in this case (see also the discussion
in the text).}
\end{center}
\end{figure}

\begin{figure}
\begin{center}
\includegraphics[width=8.6cm,height=6cm]{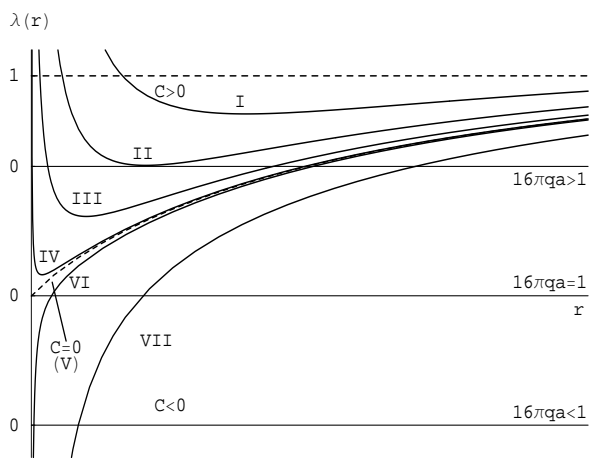}
\caption{\label{fig:6} Same comments as in Fig.4, but now for \underline{$\sigma = 1$}. The slope at the center
of the metric function for the critical configuration ($C=0$) is finite in this case (see also the discussion in the text).}
\end{center}
\end{figure}

As an illustrative example of this IRD-A2 case let us consider a one-parameter family of models supporting ESS solutions of the form

\be
E(r,Q,\mu) = \left[(r^{2}/Q) + \mu^{2}\right]^{-1/2},
\label{eq:(3-19)}
\en
where $\mu$ is the model parameter and $Q>0$ is the charge associated with the solution. Near the center these fields behave as

\be
E(r \rightarrow 0,Q,\mu) \sim \frac{1}{\mu} - \frac{r^{2}}{2Q \mu^{3}},
\label{eq:(3-20)}
\en
and we see that the maximum field strength of Eq.(\ref{eq:(2-10)}) is $a = \mu^{-1}$, the coefficient $b = (2Q \mu^{3})^{-1}$ and $\sigma = 2$. The use of the first-integral (\ref{eq:(2-6)}) leads to the expression

\be
\varphi(X,Y=0,\mu) = \frac{1}{\mu^{2}}\left[\mu^{-1} \ln\left(\frac{1 + \mu\sqrt{X}}{1 - \mu\sqrt{X}}\right) - 2\sqrt{X}\right],
\label{eq:(3-21)}
\en
for the restriction for purely electric fields of the family of Lagrangian densities supporting ESS solutions of the form (\ref{eq:(3-19)}). This expression is valid in the range $0 \leq X < 1/\mu^{2}$, where it behaves as expected for these IRD-A2 cases. The energy density for these solutions reads now as

\be
\rho(r,Q,\mu) = \frac{1}{\mu^{3}}\left[\frac{2\sqrt{1 + R^{2}}}{R^{2}} - \ln\left(\frac{\sqrt{1 + R^{2}} + 1}{\sqrt{1 + R^{2}} - 1}\right)\right],
\label{eq:(3-22)}
\en
where $R = \frac{r}{\mu \sqrt{\vert Q \vert}}$ for any sign of $Q$. The internal-energy function becomes

\bea
\varepsilon_{in}(r,Q,\mu) = \frac{8\pi \mu Q^{3/2}}{3} \Big[R\sqrt{1 + R^{2}} \nonumber \\
- R^{3} \ln \Bigg(\frac{\sqrt{1 + R^{2}} + 1}{\sqrt{1 + R^{2}} - 1} \Bigg) \\
+ 2\ln(R + \sqrt{1 + R^{2}}) \Big], \nonumber
\label{eq:(3-23)}
\ena
which exhibits the qualitative behaviour of the A2-curves of Fig.2. From these formulae the determination of the structure of the metrics and the verification of all the general statements $a)$ to $c)$ for the different subcases of this example are straightforward.

\subsection{IRD-UVD case}

In this case the internal and external integrals of energy do not exist, but the primitive of $r^{2}T_{0}^{0}(r,Q)$ in equation (\ref{eq:(3-4)}) is well defined up to an arbitrary constant

\be
\varepsilon(r,Q,\Gamma) = 4\pi \int r^{2} T_{0}^{0}(r,Q) + \Gamma.
\label{eq:(3-24)}
\en
Thus the metric is integrated as

\be
\lambda(r,Q,D) = 1 + \frac{D}{r} - \frac{2\varepsilon(r,Q,0)}{r},
\label{eq:(3-24)bis}
\en
where we have introduced the arbitrary constant $D = C-2\Gamma$. Owing to the admissibility conditions and the previous analysis, the function $r^{2} T_{0}^{0}(r,Q)$ is a monotonic function, decreasing from $+\infty$ at $r = 0$, where it behaves as $r^{p} (p \leq -1)$, and vanishing asymptotically as $r^{q} (-1 \leq q < 0)$ (see Eqs.(\ref{eq:(2-11)})) and (\ref{eq:(2-12)})). Consequently, the curves $\varepsilon(r,Q,\Gamma)$ are monotonic and convex, increasing from $-\infty$ at $r =0$ and exhibiting a horizontal parabolic branch as $r \rightarrow \infty$ (for any $\Gamma$ and, in particular, for $\Gamma=0$). The condition for the vanishing of the metric function is now

\be
\varepsilon(r,Q,\Gamma=0) = \frac{r + D}{2},
\label{eq:(3-25)}
\en
whose solutions define the radii of the horizons of the different black hole configurations (see Fig.2). There is always a value of the constant $D$ in the r.h.s. of Eq.(\ref{eq:(3-25)}) determining the straight line of the beam $\frac{r+D}{2}$ which is tangent to the energy function and defining an extreme black hole, whose associated radius $r_{hextr}(Q)$ is the solution of

\be
4\pi r^{2} T_{0}^{0}(r,Q) = \frac{1}{2},
\label{eq:(3-26)}
\en
(note that, as in the previous cases, this radius does not depend on the value of the constant $D = C -2\Gamma$). Equation (\ref{eq:(3-25)}) allows for the determination of a value of the constant $D_{extr}(Q)$ corresponding to this tangent line. We see that, aside from \textbf{extreme black holes}, there are now configurations with a \textbf{naked singularity} at the center (when $D > D_{extr}(Q)$) and also \textbf{two-horizons black holes} (when $D < D_{extr}(Q)$). Excepting by the anomalous asymptotic behaviour, these structures are qualitatively similar to those associated to the RN solutions.

The dominating term in the expression of the metric function near the center and as $r\rightarrow \infty$ is the last one of Eq.(\ref{eq:(3-24)bis}) and, consequently, this function diverges to $+\infty$ at $r=0$ and behaves asymptotically as $1-\lambda(r \rightarrow \infty, Q,D)  \sim r^{q} \rightarrow 0^{+}$ (no matter the sign of the constant $D$), exhibiting a similar form as that of the curves I to IV in Fig.3, with a single minimum $\lambda(r_{min},Q,D) < 1$. As can be easily checked by deriving Eq.(\ref{eq:(3-24)bis}) and taking into account the admissibility conditions, the values of these minima decrease without limit and their locations approach the origin as $D \rightarrow -\infty$.

As an illustrative example of this class of models let us consider the field

\be
E(r) = \frac{\sqrt{Q}}{r},
\label{eq:(3-27)}
\en
which is the ESS solution of a very simple IRD-UVD model, but exhibiting all the features of this case. The restriction to the half axis $X>0, Y=0$ of the corresponding family of Lagrangian densities supporting this solution is obtained from Eq.(\ref{eq:(2-6)}) and takes the form

\be
\varphi(X,Y=0) = \frac{2}{3} X^{3/2}.
\label{eq:(3-28)}
\en
The energy density reads

\be
T_{0}^{0}(r,Q) = \frac{4Q^{3/2}}{3r^{3}},
\label{eq:(3-29)}
\en
and the primitive of $r^{2} T_{0}^{0}(r,Q)$ is

\be
\varepsilon(r,Q) = \frac{16 \pi Q^{3/2}}{3} \ln(r)+\Gamma.
\label{eq:(3-30)}
\en
The metric function takes the form

\be
\lambda(r,D,Q) = 1 + \frac{D}{r} - \frac{32 \pi Q^{3/2}}{3} \frac{\ln(r)}{r},
\label{eq:(3-31)}
\en
which diverges to $+\infty$ at $r = 0$, reaches a minimum with $\lambda < 1$ and approaches asymptotically to $\lambda \rightarrow 1^{-}$ as $r \rightarrow \infty$, slower than the Schwarzschild solution. The extreme black hole horizon radius is

\be
r_{hextr}(Q) = \frac{32 \pi Q^{3/2}}{3},
\label{eq:(3-32)}
\en
and the corresponding value of the constant $D_{extr}(Q)$ is

\be
D_{extr}(Q) = \frac{32 \pi Q^{3/2}}{3} \left[\ln\left(\frac{32 \pi Q^{3/2}}{3}\right) - 1\right].
\label{eq:(3-33)}
\en
As easily verified, the solutions (\ref{eq:(3-31)}) with $D > D_{extr}(Q)$ have minima $\lambda_{min} > 0$ and are naked singularities. Those with $D < D_{extr}(Q)$ have $\lambda_{min} < 0$, cutting twofold the $\lambda = 0$ axis, and correspond to two-horizons black hole configurations (see Fig.3, curves I to IV).

The lagrangian (\ref{eq:(3-28)}) and the associated solutions (\ref{eq:(3-27)}) correspond, respectively, to the limit case of the IRD-A1 family (\ref{eq:(3-21)}) and their associated solutions (\ref{eq:(3-19)}) as the parameter $\mu \rightarrow 0$. In this limit the maximum field strength ($a = \mu^{-1}$) of the ESS solutions of this family diverges and the limit model becomes IRD-UVD.

\section{Summary and perspectives}

Let us summarize the main results obtained until now.

The contents of this paper, together with those of Ref.\cite{dr10}, complete the analysis of the geometrical structure of the G-ESS solutions of admissible NEDs minimally coupled to gravity. The families of Lagrangian theories considered exhaust the class of the admissible ones.

These admissible G-NEDs lead always to asymptotically flat G-ESS solutions. At large distances, the metrics approach flatness as the Schwarzschild field in cases B1, B2 and B3, with a well defined ADM mass. On the other hand, they approach flatness slower than the Schwarzschild field for models with IRD asymptotic behaviour and the ADM mass is not defined in these cases. The central-field behaviour plays no role in determining the asymptotic gravitational structure of the solutions.

The geometrical structures of the G-ESS solutions at finite $r$ are dependent on the central field behaviour of the ESS solutions in flat space or, equivalently, on the behaviour of the Lagrangian densities on the boundary of their domain of definition at $Y=0$. The present analysis, together with the one of Ref.\cite{dr10}, show that the qualitative nature of these structures is independent on the asymptotic behaviours of the G-ESS solutions. For the UVD central field cases, combined with B1, B2 or B3 behaviours as $r \rightarrow \infty$, these structures are similar as those of the Reissner-Nordstr\"om solutions of the Einstein-Maxwell field equations (naked singularities, extreme black holes and two-horizons black holes). The UVD-IRD cases exhibit similar structures, but with an anomalous (non Schwarzschild-like) asymptotic behaviour. For A1 and A2 central field cases there are in addition single-horizon black holes and (in case A2) extreme and non-extreme black points, no matter the asymptotic properties (and, hence, the finite or divergent-energy character) of the ESS solutions in flat space. But when combined with IRD behaviour at large $r$, they lead also to anomalous asymptotic behaviours for the gravitational fields.

As already mentioned, all the G-ESS solutions considered here have a curvature singularity at the center, whose causal structure is dependent on the sign of the metric function there. For NEDs with Maxwellian weak-field limit (B2 cases) this is a consequence of the Bronnikov theorem \cite{bronnikov-theorem}. It is easy to extend the validity of this theorem for the admissible models belonging to the other asymptotic cases (B1, B3 and IRD) considered here. Let us summarize the argument of Bronnikov, as presented in Ref.\cite{bronnikov00}:

Owing to the diagonal character of the components of the Ricci tensor in the coordinate system (\ref{eq:(3-2)}), the quadratic curvature invariant $R_{\alpha\beta}R^{\alpha\beta}$ can be written as the sum of squares of the diagonal components of the mixed Ricci tensor

\be
R_{\alpha\beta}R^{\alpha\beta} = \sum_{\lambda} \left(R_{\lambda}^{\lambda}\right)^{2}.
\label{eq:(3-34)}
\en
Then, the finiteness of this invariant at a given point implies the finiteness of the diagonal components of the Ricci tensor there and, consequently, of those of the energy-momentum tensor (\ref{eq:(3-1)bis}), in such a way that the quantity

\be
\frac{1}{2} \left(T_{0}^{0} - T_{2}^{2}\right) = E^{2} \frac{\partial \varphi}{\partial X} \Big \vert_{Y=0},
\label{eq:(3-35)}
\en
must be finite. On the other hand, the first-integral (\ref{eq:(2-6)}) gives

\be
E \frac{\partial \varphi}{\partial X} \Big \vert_{Y=0} = \frac{Q}{r^{2}}.
\label{eq:(3-36)}
\en
At the center, the r.h.s. of (\ref{eq:(3-36)}) diverges, and the compatibility of this expression with the finiteness of (\ref{eq:(3-35)}) leads to the conditions

\be
\frac{\partial \varphi}{\partial X} \Big \vert_{X \rightarrow 0,Y = 0} \rightarrow \infty \hspace{.15cm} ; \hspace{.15cm} E(r \rightarrow 0) \rightarrow 0.
\label{eq:(3-37)}
\en
This behaviour is incompatible with models belonging to the Maxwellian cases B2, which proves the theorem in its initial version.

The behaviour (\ref{eq:(3-37)}) is also incompatible with the cases IRD and B1, for which the theorem is automatically extended. The first one of conditions (\ref{eq:(3-37)}) is compatible with the cases B3 and, at first sight, one could expect the existence of regular G-ESS solutions for some G-NEDs of this class. However, the positivity of the energy requires for the ESS solutions that $E(r \rightarrow \infty) \rightarrow 0$. This condition, aside from the second one of Eq.(\ref{eq:(3-37)}) (and the positivity of $\frac{\partial \varphi}{\partial X} \big \vert_{Y = 0}$) breaks the monotonic character of $E(r)$. As a consequence, the first-integral (\ref{eq:(3-36)}) implies the multi-branched character of $\varphi(X,Y=0)$ and the non-admissibility. This completes the extension of the Bronnikov theorem: \textit{The G-ESS solutions of any admissible NED minimally coupled to gravity exhibit a curvature singularity at the center.}

As established in reference \cite{dr09} the stability of the ESS solutions of these models in flat space is endorsed by the necessary and sufficient condition

\be
\frac{\partial \varphi}{\partial X} \geq 2X\frac{\partial^{2} \varphi}{\partial Y^{2}},
\label{eq:(3-33)bis}
\en
to be satisfied by the Lagrangian densities in the domain of definition of the solution. The question of whether ESS solutions of admissible models satisfying this stability flat-space criterion are also stable when minimally coupled to gravity or, on the contrary, the gravitational coupling has the effect of destabilizing stable configurations in flat space (requiring new conditions to be imposed for stability in the gravitational case) is currently under analysis.

In this work we have not addressed the problem concerning the thermodynamic analogies for these models. As it has been established in Ref.\cite{rasheed97} the zeroth and first laws of black hole thermodynamics can be extended to any G-NED with Maxwellian weak-field limit (B2 cases) (although the Smarr formula cannot, in general). It is easy to extend this result to cases B1 and B3 and it is possible to obtain generalized Smarr formulae for admissible G-NEDs. These and other related issues will be analyzed elsewhere.

Finally, although it lies beyond the scope of this paper, let us mention, as another important issue in this context, the study of the dynamics of solutions and wave propagation in G-NEDs. Among the first analysis involving this kind of theories, the one of Ref.\cite{plebanski70} generalizes to the gravitating case, for BI electrodynamics, the Boillat results on wave propagation of general NEDs in flat space \cite{boillat70}. More recent advances in this line of work with applications to modern problems can be found, for example, in Ref.\cite{gibbons00}.

\begin{center}
\textbf{ACKNOWLEDGMENTS} \\
\end{center}
This work has been partially supported by Spanish grants FC-08-IB08-154 and MICINN-09-FPA2009-11061.


\begin{thebibliography}{00}

\bibitem{BI} M. Born and L. Infeld, Proc. R. Soc. A \textbf{144}, 425
(1934).
\bibitem{string-BI}  E. S. Fradkin and A. A. Tseytlin, Phys. Lett. B \textbf{163}, 123 (1985); A. Abouelsaood, C. G. Callan, Jr., C. R. Nappi and S. A. Yost, Nucl.
Phys. B \textbf{280}, 599 (1987);  R. G. Leigh, Mod.
Phys. Lett. A \textbf{4}, 2767 (1989); A. A. Tseytlin, Nucl. Phys. B \textbf{501}, 41 (1997); D. Brecher, Phys. Lett. B \textbf{442}, 117 (1998).
\bibitem{gravity-BI} B. Hoffmann, Phys. Rev. \textbf{47}, 877 (1935); A. Garcia, H. Salazar and J. F. Plebanski, Nuovo Cimento Soc. Ital. Fis. A \textbf{84}, 65 (1984);
M. Demianski, Found. Phys. \textbf{16}, 187 (1986); G. W. Gibbons and D. A. Rasheed, Nucl. Phys. B \textbf{454}, 185 (1995); N. Breton, Phys. Rev. D \textbf{67}, 124004 (2003).
\bibitem{gravity-EH} H. Yajima and T. Tamaki, Phys. Rev. D \textbf{63}, 064007 (2001).
\bibitem{oliveira94} H. P. de Oliveira, Class. Quant. Grav. \textbf{11}, 1469 (1994).
\bibitem{gravity-log} H. H. Soleng, Phys. Rev. D \textbf{52}, 6178 (1995).
\bibitem{cataldo00} M. Cataldo, N. Cruz, S. del Campo and A. Garc\'ia, Phys. Lett. B \textbf{484}, 154 (2000).
\bibitem{hassaine07} M. Hassaine and C. Martinez, Phys. Rev. D \textbf{75}, 027502 (2007).
\bibitem{hassaine08} M. Hassaine and C. Martinez, Class. Quant. Grav. \textbf{25}, 195023 (2008).
\bibitem{dr07-2} J. Diaz-Alonso and D. Rubiera-Garcia, Phys. Lett. B \textbf{657} 257 (2007).
\bibitem{dr09} J. Diaz-Alonso and D. Rubiera-Garcia, Ann. Phys. \textbf{324}, 827 (2009).
\bibitem{dr10} J. Diaz-Alonso and D. Rubiera-Garcia, Phys. Rev. D \textbf{81}, 064021 (2010).
\bibitem{bronnikov-theorem} K. A. Bronnikov, V. N. Melnikov, G. N. Shikin and K. P. Staniukowicz, Ann. Phys. \textbf{118}, 84 (1979).
\bibitem{magnetic-sol} K. A. Bronnikov, Phys. Rev. D \textbf{63}, 044005 (2001).
\bibitem{regular} A. Burinskii and S. R. Hildebrandt, Phys. Rev. D \textbf{65}, 104017 (2002); I. Dymnikova, Class. Quant. Grav. \textbf{21}, 4417 (2004).
\bibitem{ab} E. Ay\'on-Beato and A. Garc\'ia,
Phys. Rev. Lett. \textbf{80}, 5056 (1998);
E. Ay\'on-Beato and A. Garc\'ia, Gen. Rel. Grav. \textbf{31}, 629 (1999);
E. Ay\'on-Beato and A. Garc\'ia, Phys. Lett. B \textbf{464}, 25 (1999).
\bibitem{bronnikov00} K. A. Bronnikov, Phys. Rev. Lett. \textbf{85}, 4641 (2000).
\bibitem{ortin} T. Ortin, \emph{Gravity and strings}, Cambridge Monographs on Mathematical Physics (Cambridge University Press, Cambridge, U. K., 2004).
\bibitem{rasheed97} D. A. Rasheed, arXiv:hep-th/9702087.
\bibitem{plebanski70} J. Pleba\'nski, \emph{Lectures on Non-linear Electrodynamics} (NORDITA, Copenhagen, 1970).
\bibitem{boillat70} G. Boillat, J. Math. Phys. \textbf{11}, 941 (1970); G. Boillat, J. Math. Phys. \textbf{11}, 1482 (1970).
\bibitem{gibbons00} G. W. Gibbons and K. Hashimoto, JHEP \textbf{0009}, 013 (2000); M. Novello, S. E. Perez Bergliaffa and J. M. Salim, Class. Quant. Grav. \textbf{17}, 3821 (2000); M. Novello, V. A. De Lorenci, J. M. Salim and R. Klippert, Phys. Rev. D \textbf{61}, 045001 (2000); G. W. Gibbons and C. A. R. Herdeiro, Phys. Rev. D \textbf{63}, 064006 (2001).

\end{thebibliography}
\end{document}